\documentclass[aps,pra,twocolumn,showpacs,floatfix]{revtex4}
\usepackage{epsfig}
\usepackage{graphicx}
\usepackage{dcolumn}
\usepackage{amsthm,amsmath}

\begin{document}

\title{Dispersion coefficients for the interaction of Cs atom with different material media}

\author{Kiranpreet Kaur$^a$,
Jasmeet Kaur$^a$, 
B. K. Sahoo$^{b}$\footnote{Email: bijaya@prl.res.in}
and
Bindiya Arora$^a$\footnote{Email: bindiya.phy@gndu.ac.in} }
\affiliation{$^a$Department of Physics, Guru Nanak Dev University, Amritsar, Punjab-143005, India}
\affiliation{$^b$Theoretical Physics Division, Physical Research Laboratory, Navrangpura, Ahmedabad-380009, India}
\date{Recieved Date; Accepted Date}

\begin{abstract}
Largely motivated by a number of applications, the dispersion ($C_3$) coefficients for the interaction of a Cs atom with different
material media such as Au (metal), Si (semiconductor) and various dielectric surfaces like SiO$_2$, SiN$_{\rm{x}}$, sapphire and 
YAG are determined using accurate values of the dynamic polarizabilities of the Cs atom obtained employing the relativistic 
coupled-cluster approach and the dynamic dielectric constants of the walls. Moreover, we also give the retardation coefficients in 
the graphical representation as functions of separation distances to describe the interaction potentials between the Cs atom with 
the above considered material media. For the easy access to the interaction potentials at a given distance of separation, 
we devise a simple working functional fitting form for the retarded coefficients in terms of two parameters that are quoted 
for each medium.

\end{abstract}
\maketitle

\section{Introduction}\label{sec1}

Atom-surface interactions are important for understanding numerous physical, chemical and biological 
processes. Owing to this, research works of Lennard-Jones ~\cite{len}, Bardeen~\cite{bardeen}, Casimir and Polder~\cite{casimir} and Lifshitz ~\cite{lifshitz} 
have drawn a lot of attention over the last few decades ~\cite{wylie1,wylie2,barut1,barut2,jent}. In the non-retarded regime (at small distances), 
the forces between the fluctuating atomic dipole and its immediate image associated with the polarization charges induced in the surface brings about 
the atom-surface van der Waals interactions~\cite{bout,len}. Thus, for the short separation distance `$a$' between the atom and surface, the interaction 
energy scales as $1/a^3$ (at the intermediate distances, the retardation effects are taken into account by introducing a damping function), while for 
large atom-surface distances as compared to a typical atomic wavelength, the interaction energy scales as $1/a^4$ ~\cite{casimir,mavro,mcl,boyer,kohn}. 

Investigations of the van der Waals dispersion forces between an atom and a surface are of immense interest to the physicists working in various 
domains of physical sciences. Assessing these forces accurately can result in new pathways towards engineering, technology and research. Finding out 
well behaved interactions of atoms or molecules with atomically well defined surfaces are beneficial for advocating future device applications at 
the nanometer dimensions~\cite{song}. These atom-surface interactions can cause shifts in the oscillation frequencies of the trap that can change the 
trapping conditions. Therefore, when the trapping potentials are determined, particularly at the magic wavelengths, these effects need to be accounted for
\cite{ant}. A comprehensive cognizance of the dispersion coefficients is necessary for the experimental studies of photoassociation, realization of 
Bose-Einstein condensates (BECs), interpreting fluorescence spectroscopy, determination of scattering lengths, and analysis of Feshbach resonances~\cite{robert,
amiot,leo,harber,lean,lin,bindiya}. Knowing their behaviors are significantly important in generating novel atom optical devices, also known as the atom 
chips~\cite{folman}. 

\begin{figure*}[t]
\includegraphics[width=17.5cm, height=8.5cm]{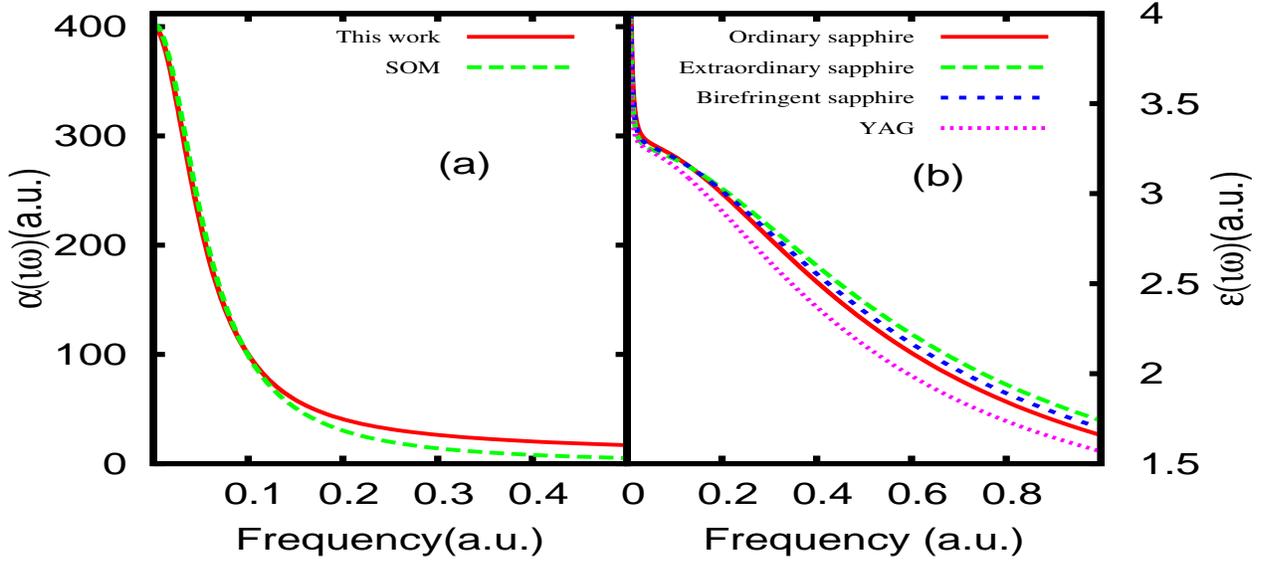}
\caption{(Color online) (a) Dynamic polarizabilities of Cs atom from the present calculations as function of frequency are shown and compared 
with the results obtained using single oscillator model (SOM), (b) Imaginary dynamic dielectric constants of various dielectric surfaces (ordinary sapphire, extraordinary sapphire, 
birefringent sapphire and YAG) along the imaginary axis as function of frequency.}~\label{dype}
\end{figure*}

The atom-wall interaction potentials have significant dependences on the dielectric properties of the materials~\cite{kharchenko,lach1,caride,babb1}. 
Majority of the earlier theoretical works involve the interactions of the ground states of atoms with metallic or dielectric 
surfaces, considering both the non-retarded and retarded van der Waals forces ~\cite{len, bardeen,casimir,lifshitz,mavro,kohn}.
Since precise measurement of the interaction potentials or in that sense the dispersion coefficients 
are extremely difficult, very few experimental investigations of the van der Waals interactions between the atoms in their ground
states and solid state surfaces have been accomplished ~\cite{vida}. However, 
very precise measurements of ratios of the dispersion coefficients of some of the alkali atoms with SiN$_{\rm{x}}$ have been 
reported \cite{lonij1,lonij2}. This gives an opportunity to test the validity of a method to reproduce the experimental results. 
For the theoretical determinations, the $C_3$ coefficients can be efficiently 
expressed in terms of the conducting properties of trapping media and the dynamic dipole polarizabilities of the interacting atom following the Lifshitz 
theory~\cite{lifshitz,lifshitzbook}.
Available theoretical studies 
on the Cs atom-wall interactions are performed considering approximated values of either the dynamic polarizabilities of the Cs atom or the conducting 
properties of the materials~\cite{cronin,lach1,caride}. In our previous work, we have demonstrated importance of using accurate values of the atomic dynamic
polarizabilities in order to compare ratios of the dispersion coefficients of the light nuclei alkali atoms with their corresponding measured values
\cite{bindiya}. It was shown that these interactions can be estimated with sufficient accuracy using the relativistic coupled-cluster (RCC) method.  

In this work, we intend to calculate the van der Waals interaction potentials between the Cs atom in the ground state with the trapping materials as an
ideal conductor, a normal good conductor like Au, a semiconductor like Si, dielectric objects like SiO$_2$, SiN$_{\rm{x}}$, sapphire and a YAG surface by using 
accurate polarizability values of the Cs atom. Combining these results with our previously estimated $C_3$ coefficients for the other alkali
atoms \cite{bindiya}, we would like to present ratios of these constants with respect to the Cs atom. Unless stated otherwise, we use atomic units (a.u.) throughout the paper.

\section{Method of Calculations}\label{sec2}

The formula given by E. M. Lifshitz and collaborators in Moscow in 1955 can efficiently be applied to describe the van der Waals and Casimir-Polder 
interactions between an atom and a semispace, a material plate, or a layered structure \cite{lifshitz,lifshitzbook}. For the intermediate 
separations, where the atom-wall force of attraction is almost negligible, the atom-wall interactions can be computed by considering a polarizable 
particle interacting with a surface or a wall as a continuous medium having a frequency dependent permittivity $\epsilon(\omega)$. The working 
formula for the interaction potential in terms of the dielectric constants is given by~\cite{lach1,lifshitz,lifshitzbook,bindiya}
\begin{eqnarray}
V(a)&=& -\frac{\alpha_{fs}^3}{2\pi}\int_0^{\infty}d\omega \omega^3\alpha(\iota\omega) \nonumber \\ 
&& \times \int_1^{\infty}d\xi e^{-2\alpha_{fs}\xi\omega a} H(\xi,\epsilon(\iota\omega)), \label{atwp}
\end{eqnarray}
where $\alpha_{fs}$ is the fine structure constant, $\alpha(\iota \omega)$ is the dynamic dipole polarizability of the atom, $a$ is the
distance of separation between the Cs atom and a surface or a wall and $\epsilon(\iota \omega)$ is the frequency dependent dielectric constant of
the wall. The expression for $H(\xi, \epsilon(\iota \omega))$, which is a function of the dielectric permittivity of the material wall, is given by
\begin{equation}
H(\xi,\epsilon(\iota \omega))=(1-2\xi^2)\frac{\xi^{'}-\epsilon\xi}{\xi^{'}+\epsilon\xi} + \frac{\xi^{'}-\xi}{\xi^{'}+\xi},
\end{equation}
where $\xi^{'}=\sqrt{\xi^2+\epsilon-1}$. Evaluation procedure of this functional form is described in Refs.~\cite{lach1,bindiya}.
A more general expression for the potential as described in Eq.(\ref{atwp}) for both the retarded and short distances is conveniently expressed by
\begin{equation}
V(a)=-\frac{C_3}{a^3} f_{3}(a), \label{veq}
\end{equation}
 where the expression for the dispersion coefficient $C_3$ is given by
\begin{equation}
C_3=-\frac{1}{4\pi }\int_0^\infty d \omega \alpha(\iota\omega) S(\iota\omega), \label{c3eq}
\end{equation}
for
\begin{eqnarray}
 S(\iota \omega) = \frac{\epsilon(\iota\omega)-1}{\epsilon(\iota\omega)+1} \label{seq}.
\end{eqnarray}
 At the short 
distance, this interaction potential can be approximated to a simple form as~\cite{lifshitz,fichet}
\begin{equation} 
 V(a)=-\frac{C_3}{a^3}.  \label{short} 
 \end{equation}
For a perfect conductor, $\epsilon(\omega)\rightarrow \infty$. Thus, it yields
\begin{equation}
C_3=-\frac{1}{4\pi }\int_0^\infty d \omega \alpha(\iota\omega) \label{c3p}
\end{equation}
and
\begin{equation}
f_{3}(a)=\frac{1}{4\pi C_3}\int_0^{\infty}d\omega \alpha(\iota \omega)e^{-2\alpha_{fs}\omega a} P^{(\infty)}(\alpha_{fs} \omega a) \label{f3p}
\end{equation}
with $P^{(\infty)}(x)=1+2x+2x^{2}$.

Adopting a similar approach as was done in Ref.~\cite{bindiya}, we actuate to evaluate the $C_3$ coefficients and $f_{3}(a)$. We 
calculate the $C_3$ coefficients using Eq. (\ref{c3eq}) and interaction potentials using Eq. (\ref{atwp}). By combining these 
two quantities, we determine the retarded function $f_3(a)$ using Eq.(\ref{veq}). To calculate the $C_3$ and retarded functions for a perfect conductor, we use Eqs. 
(\ref{c3p}) and (\ref{f3p}) directly. As can be noticed, accurate evaluation of these quantities entirely depend on the accuracies 
in the values of the dynamic polarizabilities of the atom and dynamic dielectric constants of the material or a wall. Thus, 
importance of the present work lies in the rigorous determination of these dynamic properties of the Cs atom and materials under 
consideration. We describe below the approach considered to evaluate the dynamic polarizabilities of the Cs atom and dynamic 
dielectric constants of Au, Si, and SiO$_2$, SiN$_{\rm{x}}$, sapphire and YAG surfaces having conducting, semiconducting and dielectric 
characteristics, respectively.

\begin{table}
\caption{\label{pol} The E1 matrix elements and various contributions to the scalar polarizability of the ground state in Cs atom.}
\begin{ruledtabular}
\begin{tabular}{lcc}
Transition	& E1 amplitude (a.u.)		& $\alpha$ (a.u.)	\\
\hline
& & \\
6$s_{1/2}$-6$p_{1/2}$		& 4.489~\cite{rafac}		& 131.88	\\
6$s_{1/2}$-7$p_{1/2}$		& 0.32				& 0.34		\\
6$s_{1/2}$-8$p_{1/2}$		& 0.10				& 0.03		\\
6$s_{1/2}$-9$p_{1/2}$		& 0.05				& 0.01		\\
6$s_{1/2}$-10$p_{1/2}$		& 0.03				& $\sim0$		\\
6$s_{1/2}$-6$p_{3/2}$		& 6.324~\cite{rafac}		& 249.38	\\
6$s_{1/2}$-7$p_{3/2}$		& 0.64				& 1.37		\\
6$s_{1/2}$-8$p_{3/2}$		& 0.25				& 0.18		\\
6$s_{1/2}$-9$p_{3/2}$		& 0.14				& 0.05		\\
6$s_{1/2}$-10$p_{3/2}$		& 0.09				& 0.02		\\
\\
$\alpha_{v}$		&		& 383.30		\\
$\alpha_{c}$		&		& 16.8			\\
$\alpha_{cv}$		&		& $-0.5$			\\
$\alpha_{t}$		&		& 0.2			\\
Total 	&	& 399.8			\\
 & & \\
 Others	&	& 399~\cite{borch},399.9~\cite{dere3}	\\
 Experiment	&	& 401.0(6)~\cite{amini}
\end{tabular} 
\end{ruledtabular}
\end{table}

\begin{table}
\caption{\label{c3} Calculated $C_3$ coefficients for the interaction of the Cs atom with the perfect conductor, Au (metal), Si (semiconductor) 
and the dielectric surfaces (SiO$_2$, SiN$_{\rm{x}}$, Sapphire and YAG) along with the classification of contributions from various parts of 
the dynamic polarizabilities.}
\begin{ruledtabular}
\begin{tabular}{lccccc}
     		&  Core       &  Valence          & Core-        & Tail            & Total                \\
     		&	      &			   & Valence	   &		     &			     \\
\hline
& & \\
Ideal conductor  & 2.350     & 2.5309     	&$-0.043$      & 0.004         & 4.8427    		\\
		    &		&		&		&		& 4.268~\cite{dere3}		\\
Metal: Au    & 0.706     &  2.191     	& $-0.017$     & 0.003    & 2.8823       \\
	     &		  &		&		&	   & 2.79~\cite{lach1}		\\
& & \\
Semiconductor: Si   & 0.512  & 1.874    &$-0.0131$   & 0.0025    & 2.3756    \\
& & \\
Dielectric:	&		&		&		&		&	\\
SiO$_2$     & 0.310  & 0.881    &$-0.0077$   & 0.0012    & 1.1846    \\
SiN$_{\rm{x}}$     & 0.383  & 1.335    &$-0.0098$   & 0.0018    & 1.7100    \\
Ordinary   & 0.527  & 1.319    &$-0.0127$   & 0.0019    & 1.8360    \\
sapphire                 &	&	&	&	&	   \\
Extraordinary    & 0.551  & 1.315    &$-0.0132$   & 0.0019    & 1.8542    \\
sapphire			&	&	&	&	&	   \\
Birefringent      & 0.5391  & 1.317    &$-0.0129$   & 0.0019    & 1.84523    \\
sapphire		&	&	&	&	&	   \\
YAG         & 0.490  & 1.283    &$-0.01975$   & 0.0018    & 1.7635  
\end{tabular} 
\end{ruledtabular}
\end{table}
The procedure for determining accurate values of the dynamic polarizability of an atomic system having a closed core and a valence 
electron is given by us in Ref.~\cite{nandy,jasmeet}. We apply the same procedure here to calculate the dynamic polarizabilities of the ground 
state of Cs. In this approach, we divide contributions to polarizability $\alpha$ into three parts as~\cite{nandy,jasmeet}
\begin{equation}
\alpha=\alpha_{v}+\alpha_{c}+\alpha_{cv} \nonumber 
\end{equation}
where $\alpha_v$, $\alpha_c$ and $\alpha_{cv}$ correspond to the polarizabilities due to the correlations due to the valence 
electron, core electrons and core-valence electrons correlations to the polarizability, respectively. It is known that major 
contributions to the alkali atoms come from $\alpha_v$ \cite{dere1,dere2,bindiya,bindiya1,bindiya2}. We evaluate this 
contribution by considering predominantly contributing electric dipole (E1) matrix elements between the ground state and many 
low-lying excited states of Cs in a sum-over-states approach combining with the experimental energies. We use the precisely values of the 
E1 matrix elements for the predominantly contributing low-lying transitions estimating from the precisely measured lifetimes of the 
$6p ~{^2}P_{1/2}$ and $6p ~{^2}P_{3/2}$ states \cite{rafac}. The other important E1 matrix elements are obtained using the RCC method in the 
singles and doubles approximation (CCSD method) as described in \cite{bijaya1,bijaya2}. 
Contributions to $\alpha_c$ are determined using a relativistic random-phase approximation (RRPA) as described in 
\cite{jasmeet}. It has been demonstrated that the RRPA method can give rise to very reliable results for the atomic systems having 
inert gas configurations ~\cite{yashpal}. Smaller contributions from $\alpha_{cv}$ and from the high-lying excited 
states (tail contribution $\alpha_t$) that are omitted in the above sum-over-states approach are estimated in the Dirac-Hartree-Fock 
(DHF) approximation.

It is not easy to get the dynamic electric permittivity of any real materials, so the convenient way of determining these constants 
for simple metals such as gold is to use the Drude-Lorentz model as was done in Ref.~\cite{lach1}. In pursuance of obtaining more 
realistic values of these constants for different materials, we prefer to use the known real and imaginary parts of the refractive 
indices of a material at few real values of frequency $\omega$. The imaginary parts of the dielectric permittivities of the materials 
can then be obtained using the relation
\begin{equation}
Im \left[\epsilon(\omega)\right]=2n(\omega)\kappa(\omega), 
\end{equation}
where $n(\omega)$ and $\kappa(\omega)$ are the respective real and imaginary parts of the refractive index of a material at 
frequency $\omega$. We use the optical data from the handbook by Palik ~\cite{palik} for the frequencies ranging from 0.1 eV 
to 10000 eV for Au metal to calculate $Im [\epsilon(\omega)]$. Thereafter, the required real values of the dielectric constants at the imaginary frequencies ($\epsilon(\iota\omega)$) are obtained by using the Kramers-Kronig formula. The available data, however, does not cover the whole frequency 
range to carry out the integration of Eq. (\ref{c3eq}). Thus, we extrapolate these values for the lower 
frequencies to increase the domain over which the integrations are to be performed ~\cite{lamb1,klim}. For the frequencies below 
0.1 eV, the classified values from ~\cite{palik} are extrapolated using the free electron Drude model in which the dielectric 
permittivity along the imaginary axis is represented as
\begin{equation}
\epsilon(\iota\omega)=1-\frac{{\omega_p}^2}{\omega(\omega+\iota\gamma)} ,
\end{equation}
where $\omega_p=\left(2\pi c/\lambda_p\right)$ is the plasma frequency and $\gamma$ is the relaxation frequency. The optical data 
values for $\omega_p$ and $\gamma$ available from various sources differ slightly, but we use these values as $\omega_p=9.0$ 
eV and $\gamma=0.035$ eV as outlined in ~\cite{lamb1,lach1,lamb,klim}. In case of Si (semiconductor),
SiO$_2$ (dielectric), Sapphire and YAG, the complex frequency-dependent dielectric permittivity are quoted for a wide range of 
energies in the handbook of Palik. Therefore, we use all these values for carrying out the integration and do not extrapolate any 
data. On the otherhand, experimental data of $n(\omega)$ and $k(\omega)$ for SiN$_{\rm{x}}$ are not available at all and we use 
Tauc-Lorentz model~\cite{cronin,link} for estimating dielectric constants of this material.  

Again, the interactions between the ground states of atoms with an anisotropic surface have been studied before~\cite{kiraha,bruch}. 
These studies demonstrate that for a uniform birefringent dielectric surface, with the symmetry axis normal to the interface, the 
interaction potential described Eq. (\ref{veq}) is still applicable if $\epsilon$ is replaced by an effective quantity 
$\bar{\epsilon}$ defined as
\begin{equation}
\bar{\epsilon}(\iota\omega)={\left[\epsilon_{||}(\iota\omega)\epsilon_{\bot}(\iota\omega)\right]}^{\frac{1}{2}}, 
\end{equation}
where $\epsilon_{||}$ and $\epsilon_{\bot}$ are the respective dielectric permittivities for the electric fields parallel and 
perpendicular to the interface between the atom and the dielectric surface.

\begin{table}
\caption{\label{f3t} Comparison of the calculated values of damping function $f_{3}(a)$ defined in Eq.(\ref{c3eq}) for different separation distances
$a$ for the interaction of Cs atom with the gold surface.}
\begin{ruledtabular}
\begin{tabular}{lcc}
  $a$               &\multicolumn{2}{c}{$f_3(a)$}		\\
(a. u.)		& This work		& Ref.~\cite{lach1}		\\
\hline
& & \\
1$\times10^{1}$		& 1.00		& 0.99780		\\
2$\times10^{1}$		& 1.00		& 0.99408		\\
5$\times10^{1}$		& 0.98		& 0.97903		\\
1$\times10^{2}$		& 0.95		& 0.95169		\\
2$\times10^{2}$		& 0.89		& 0.90210		\\
5$\times10^{2}$		& 0.78		& 0.79521		\\
1$\times10^{3}$		& 0.66		& 0.68309		\\
2$\times10^{3}$		& 0.53		& 0.54485		\\
5$\times10^{3}$		& 0.33		& 0.33918		\\
\end{tabular}
\end{ruledtabular}
\end{table}

\begin{table}
\caption{\label{ratios}Comparison of $C_3$ coefficients ratios of Li, Na, K and Rb atoms with the Cs atom for the perfect conductor, metal, 
semiconductor and dielectric surfaces.}
\begin{ruledtabular}
\begin{tabular}{lcccc}
&  $C_3^{\text{Li}}/C_3^{\text{Cs}}$  &  $C_3^{\text{Na}}/C_3^{\text{Cs}}$  & $C_3^{\text{K}}/C_3^{\text{Cs}}$   & $C_3^{\text{Rb}}/C_3^{\text{Cs}}$  \\
\hline
& & \\
Perfect conductor	& 0.313		& 0.393		& 0.638		& 0.773		\\
Metal: Au			& 0.416		& 0.464		& 0.715		& 0.811		\\
Semiconductor: Si		& 0.430		& 0.477		& 0.726		& 0.818		\\
Dielectric:		& 	&	&	&		\\
SiO$_2$		& 0.409		& 0.458		& 0.708		& 0.807		\\
SiN$_{\rm{x}}$		& 0.418		& 0.464		& 0.709		& 0.801		\\
\end{tabular}
 
\end{ruledtabular}
 
\end{table}
\section{Results and Discussion}\label{sec3}

The evaluation of $C_3$ coefficients requires the precise estimation of dynamic polarizabilities of the Cs atom. Table~\ref{pol} presents the
scalar polarizabilities of the Cs atom in its ground state along with the E1 matrix elements for different transitions that are used to
estimate $\alpha_v$ and other contributions to the polarizabilities. The E1 matrix elements for the 6$S_{1/2}$-6$P_{1/2,3/2}$ transitions are taken from the experimentally measured
data given in ~\cite{rafac}. Our calculated value of the scalar polarizability ($\alpha(0)$) for the 6$S$ state is 399.8 a.u.. This is in very 
good agreement with the value (399 a.u.) obtained by Borschevsky \textit{et al.}~\cite{borch} and experimentally measured value 401.0(6) a.u. 
of Amini \textit{et al.}~\cite{amini}. As seen, contributions from $\alpha_{cv}$ and $\alpha_t$ are quite small and justify use of a lower 
method for their evaluation. Our $\alpha_c$ from RRPA also match with the value of Derevianko \textit{et al.} \cite{dere2}. Good agreement between the 
calculated and experimental results of $\alpha(0)$ indicates that our approach can also give similar accuracies for the estimated dynamic 
polarizabilities. 
\begin{figure}[t]
\includegraphics[scale=0.68]{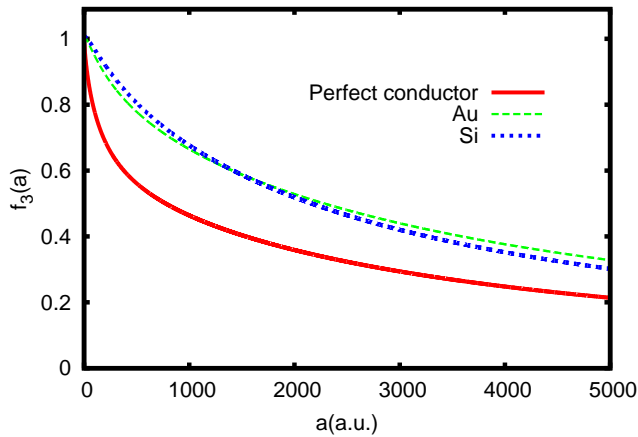}
\caption{(Color online) The retardation coefficient $f_3(a)$ for the Cs atom as a function of the distance $a$ from a perfect conductor, Au (metal) 
metal and Si (semiconductor).}~\label{csm1}
\end{figure}
\begin{figure}[t]
\includegraphics[scale=0.68]{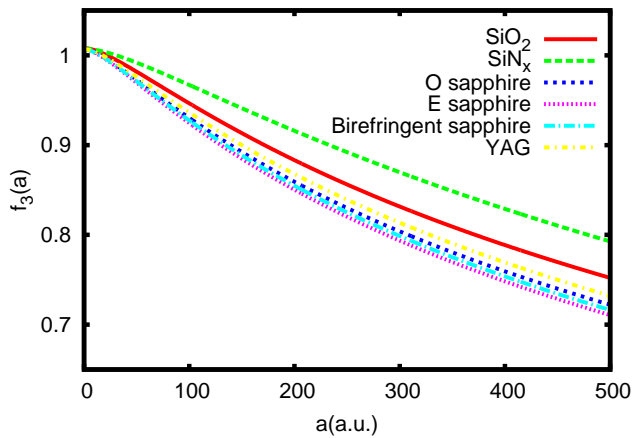}
\caption{(Color online) The retardation coefficient $f_3(a)$ for the Cs atom as a function of the distance $a$ from the different dielectric 
surfaces such as SiO$_2$, SiN$_{\rm{x}}$, Sapphire (ordinary, extraordinary, birefringent) and YAG.}~\label{csm2}
\end{figure}
Next, we compute the atom-surface dispersion $C_3$ coefficients for a perfect conductor by numerically evaluating Eq. (\ref{c3eq}).
Dynamic polarizabilities used for this purpose are plotted in Fig. \ref{dype}(a). In the same plot, we also show the values obtained by the
single oscillator model (SOM) that are used by other works as discussed below. We compare our $C_3$ coefficient for the perfect conductor 
with the result obtained by Derevianko and co-workers~\cite{dere2} in Table \ref{c3}. Though our $\alpha(0)$ value match quite well 
with Derevianko \textit{et al.}~\cite{dere3}, but we find difference in the $C_3$ value. A similar finding was also observed for the other alkali atoms that 
were studied by us in Ref. \cite{bindiya}. Our analysis shows that the main reason for the discrepancy is because of different numerical integration 
methods used in both the works. Derevianko \textit{et al.} use a Gaussian quadrature integration 
method with 50 point formula while we have used an exponential grid in our calculations as discussed 
in ~\cite{bindiya}. In the same table we have given our calculated $C_3$ coefficients for the other material media using their dynamic dielectric 
constants that are plotted in Fig. \ref{dype}(b). In our earlier work \cite{bindiya}, we have also given these constants for Au, Si, SiO$_2$ 
and SiN$_{x}$. Hence, we only present the final results without giving the fine details of their evaluations. Compared to these materials we find a different trend of $\epsilon(\iota \omega)$ values
for the ordinary, extraordinary and birefringent sapphires and also for the YAG surfaces, especially at the low frequency range. We could not find
another study to verify directly the validity of this trend, but a recent calculation of $S(\iota \omega)$ in Ref.~\cite{lali}
shows almost a similar trend. This somewhat assures us about accurate determination of the $\epsilon(\iota \omega)$ 
values for the sapphires and YAG surfaces.  

We also give the $C_3$ coefficients for all the considered material media in Table \ref{c3} along with the values known in the literature. 
As seen, there is another evaluation of $C_3$ for the Au metal reported by Lach \textit{et al.} ~\cite{lach1} They use the SOM model to estimate the dynamic polarizability values whereas the dynamic dielectric constants are estimated using the Drude model. 
Nevertheless, we find a reasonable agreement between these results. We, however, did not find any data to compare our results for the dielectric 
materials and semiconductors. From Table \ref{c3}, it can be seen that the valence correlation is dominant among the core, core-valence and tail 
correlations, but core contributions are quite significant as compared to their contributions in the evaluation of the polarizabilities. 
Among all the interacting media, the $C_3$ coefficients for the interacting perfect conductor is the highest and it is 
approximately 40\%, 51\%, 75\%, 64\%, 62\% and 63\% larger than the Au, Si, SiO$_2$, SiN$_{\rm{x}}$, sapphire and YAG surfaces respectively. The 
decrease in the interaction coefficients for the cases of dielectric media might be due to the charge dangling bonds in the materials which 
accounts for the additional interactions in the dielectrics at the shorter separations ~\cite{foster}.

Using our described procedure, we also evaluate the retarded $f_3$ functions for all the considered materials interacting with Cs. In Table 
\ref{f3t}, we compare our results for $f_3$ for the interaction of the Cs atom with the Au surface with the results obtained by Lach \textit{et al.} 
~\cite{lach1} at certain separation distances. We also find reasonable agreement between both the results. It can also be observed from this 
table that at the short separation distances ($a\rightarrow0$), the retardation coefficient $f_3(a)\rightarrow 1$. This implies that our 
calculations follow the right trend and it justifies that the expression for the van der Waals interaction potential given in Eq.(~\ref{veq}) 
reduces to Eq.(~\ref{short}) as was analytically argued in Ref. \cite{lach1}.

We have also given ratios of $C_3$ constants in Table \ref{ratios} between the Cs atom with the other alkali atoms that were reported by us 
earlier \cite{bindiya}. We observe that ratios for the $C_3$ coefficients of any of these atoms with Cs are approximately comparable 
irrespective of the surface except for the case of a perfect conductor which shows a slight deviation. We compare our results of dielectrics 
for the surfaces of SiO$_2$ and SiN$_{\rm{x}}$ only as there is not very much difference in the $C_3$ coefficient values of these surfaces with those 
of sapphire (for all ordinary, extraordinary and birefringent) and YAG.

In Fig.~\ref{csm1}, we show a comparison of the $f_{3}(a)$ values calculated for the Cs atom as a function of the atom-wall separation distance 
$a$ considering three material mediums such as a perfect conductor, Au (metal) and a Si (semiconductor). 
The retardation effects associated with the Au and Si surfaces are nearly identical at the short and intermediate separations while shows slight
deviation at large distances of separation. As expected, the $f_3$ coefficients for the perfect conductor are found to be the smallest and are 
approximately 31\% less than those of Au and Si at around 1200 a.u.. Similar plot is drawn in Fig.~\ref{csm2} to show the interaction potential 
of the Cs atom with dielectric surfaces such as SiO$_2$, SiN$_{\rm{x}}$, ordinary sapphire, extraordinary sapphire, birefringent sapphire and 
YAG. The $C_3$ coefficients of the considered dielectric surfaces are approximately comparable and hence, their $f_3$ values cannot be clearly 
distinguished when viewed over a large regime of separation distances. We, however, choose a range of separation distance up to $500$ nm in this 
plot. As can be seen from the figures that, among the various dielectric surfaces, the retardation coefficients are strongest for SiN$_{\rm{x}}$. 

\begin{table}
\caption{\label{fit} Fitting parameters $a$ and $b$ for the evaluation of the $f_3$ coefficients with the perfectly conducting wall, Au, Si, SiO$_2$,
SiN$_{\rm{x}}$, ordinary sapphire, extraordinary sapphire, birefringent sapphire and YAG surfaces.} 
\begin{ruledtabular}
\begin{tabular}{lcccc}
Surface		&	& $B1$	&	& $B2$			\\
\hline 
& & \\
Perfect Conductor	&	& 1.3077	&	& 0.1011	\\
Metal: Au		&	& 1.0563	&	& 0.0569	\\
Semiconductor: Si	&	& 1.0013	&	& 0.0636	\\
Dielectric:		&	&		&	&		\\
SiO$_2$			&	& 1.0453	&	& 0.0714	\\
SiN$_{\rm{x}}$			&	& 0.9977	&	& 0.0699	\\
Ordinary Sapphire	&	& 1.0663	&	& 0.0781	\\
Extraordinary Sapphire 	&	& 1.0741	&	& 0.0809	\\
Birefringent Sapphire	&	& 1.0702	&	& 0.0777	\\
YAG			&	& 1.0535	&	& 0.0776	\\
\end{tabular}
\end{ruledtabular}
\end{table}
In order to generate our results for the $f_3$ coefficients at a given distance of separation for future theoretical and experimental
verifications or for various applications, we devise a logistic functional form for the retardation coefficient as a function of separation 
distance as
\begin{equation}
f_{3}(a)=\frac{1}{B1+B2(\alpha_{fs}a)} .
\end{equation}
By fitting the above resulting values of $f_3$ as function of separation distance for different interacting media using the above functional 
form, we obtain the fitting parameters $A$ and $B$, which are tabulated in Table \ref{fit}. For the practical purpose, these parameters can be
used to get the $f_3$ data at a given distance.

\section{Conclusion}

In the present work, we have analyzed the atom-surface interactions between the cesium atom in its ground state with a perfect conductor, Au
(metal), Si (semiconductor) and various dielectric surfaces (SiO$_2$, SiN$_{\rm{x}}$, sapphire and YAG). We have calculated the dispersion $C_3$ 
coefficients for the above interactions and then studied the retardation effects of the above interactions by plotting the damping coefficient 
$f_3$ values as function of separation distances. More accurate dynamic polarizabilities of Cs atom are used in determining these coefficients. 
Furthermore, readily usable functional form of the retardation coefficients for the above atom-surface interactions is devised and the fitting
parameters are given.

\section*{Acknowledgement}

The work of B.A. is supported by CSIR grant no. 03(1268)/13/EMR-II, India. K.K. acknowledges the financial support 
from DST (letter no. DST/INSPIRE Fellowship/2013/758). J.K. gratefully acknowledges UGC-BSR (Grant No. F.7-273/2009/BSR) for
funding. B.K.S. acknowledges use of Vikram-100 HPC cluster in Physical Research Laboratory.


\begin{thebibliography}{56}
\expandafter\ifx\csname natexlab\endcsname\relax\def\natexlab#1{#1}\fi
\expandafter\ifx\csname bibnamefont\endcsname\relax
  \def\bibnamefont#1{#1}\fi
\expandafter\ifx\csname bibfnamefont\endcsname\relax
  \def\bibfnamefont#1{#1}\fi
\expandafter\ifx\csname citenamefont\endcsname\relax
  \def\citenamefont#1{#1}\fi
\expandafter\ifx\csname url\endcsname\relax
  \def\url#1{\texttt{#1}}\fi
\expandafter\ifx\csname urlprefix\endcsname\relax\def\urlprefix{URL }\fi
\providecommand{\bibinfo}[2]{#2}
\providecommand{\eprint}[2][]{\url{#2}}

\bibitem[{\citenamefont{Lennard-Jones}(1932)}]{len}
\bibinfo{author}{\bibfnamefont{J.~E.} \bibnamefont{Lennard-Jones}},
  \bibinfo{journal}{Trans. Faraday Soc.} \textbf{\bibinfo{volume}{28}},
  \bibinfo{pages}{333} (\bibinfo{year}{1932}).

\bibitem[{\citenamefont{Bardeen}(1940)}]{bardeen}
\bibinfo{author}{\bibfnamefont{J.}~\bibnamefont{Bardeen}},
  \bibinfo{journal}{Phys. Rev.} \textbf{\bibinfo{volume}{58}},
  \bibinfo{pages}{727} (\bibinfo{year}{1940}).

\bibitem[{\citenamefont{Casimir and Polder}(1948)}]{casimir}
\bibinfo{author}{\bibfnamefont{H.~B.~G.} \bibnamefont{Casimir}}
  \bibnamefont{and} \bibinfo{author}{\bibfnamefont{D.}~\bibnamefont{Polder}},
  \bibinfo{journal}{Phys. Rev.} \textbf{\bibinfo{volume}{73}},
  \bibinfo{pages}{360} (\bibinfo{year}{1948}).

\bibitem[{\citenamefont{Lifshitz}(1955)}]{lifshitz}
\bibinfo{author}{\bibfnamefont{E.~M.} \bibnamefont{Lifshitz}},
  \bibinfo{journal}{Zh. Eksp. Teor. Fiz.} \textbf{\bibinfo{volume}{29}},
  \bibinfo{pages}{94} (\bibinfo{year}{1955}).

\bibitem[{\citenamefont{Wylie and Sipe}(1984)}]{wylie1}
\bibinfo{author}{\bibfnamefont{J.~M.} \bibnamefont{Wylie}} \bibnamefont{and}
  \bibinfo{author}{\bibfnamefont{J.~E.} \bibnamefont{Sipe}},
  \bibinfo{journal}{Phys. Rev. A} \textbf{\bibinfo{volume}{30}},
  \bibinfo{pages}{1185} (\bibinfo{year}{1984}).

\bibitem[{\citenamefont{Wylie and Sipe}(1985)}]{wylie2}
\bibinfo{author}{\bibfnamefont{J.~M.} \bibnamefont{Wylie}} \bibnamefont{and}
  \bibinfo{author}{\bibfnamefont{J.~E.} \bibnamefont{Sipe}},
  \bibinfo{journal}{Phys. Rev. A} \textbf{\bibinfo{volume}{32}},
  \bibinfo{pages}{2030} (\bibinfo{year}{1985}).

\bibitem[{\citenamefont{Barut and Dowling}(1987)}]{barut1}
\bibinfo{author}{\bibfnamefont{A.~O.} \bibnamefont{Barut}} \bibnamefont{and}
  \bibinfo{author}{\bibfnamefont{J.~P.} \bibnamefont{Dowling}},
  \bibinfo{journal}{Phys. Rev. A} \textbf{\bibinfo{volume}{36}},
  \bibinfo{pages}{2550} (\bibinfo{year}{1987}).

\bibitem[{\citenamefont{Barut and Dowling}(1990)}]{barut2}
\bibinfo{author}{\bibfnamefont{A.~O.} \bibnamefont{Barut}} \bibnamefont{and}
  \bibinfo{author}{\bibfnamefont{J.~P.} \bibnamefont{Dowling}},
  \bibinfo{journal}{Phys. Rev. A} \textbf{\bibinfo{volume}{41}},
  \bibinfo{pages}{2284} (\bibinfo{year}{1990}).

\bibitem[{\citenamefont{Jentschura}(2015)}]{jent}
\bibinfo{author}{\bibfnamefont{U.~D.} \bibnamefont{Jentschura}},
  \bibinfo{journal}{Phys. Rev. A} \textbf{\bibinfo{volume}{91}},
  \bibinfo{pages}{010502(R)} (\bibinfo{year}{2015}).

\bibitem[{\citenamefont{Boustimi et~al.}(2001)\citenamefont{Boustimi,
  ViarisdeLesegno, Baudon, Robert, and Ducloy}}]{bout}
\bibinfo{author}{\bibfnamefont{M.}~\bibnamefont{Boustimi}},
  \bibinfo{author}{\bibfnamefont{B.}~\bibnamefont{ViarisdeLesegno}},
  \bibinfo{author}{\bibfnamefont{J.}~\bibnamefont{Baudon}},
  \bibinfo{author}{\bibfnamefont{J.}~\bibnamefont{Robert}}, \bibnamefont{and}
  \bibinfo{author}{\bibfnamefont{M.}~\bibnamefont{Ducloy}},
  \bibinfo{journal}{Phys. Rev. Lett.} \textbf{\bibinfo{volume}{86}},
  \bibinfo{pages}{2766} (\bibinfo{year}{2001}).

\bibitem[{\citenamefont{Mavroyannis}(1963)}]{mavro}
\bibinfo{author}{\bibfnamefont{C.}~\bibnamefont{Mavroyannis}},
  \bibinfo{journal}{Mol. Phys.} \textbf{\bibinfo{volume}{6}},
  \bibinfo{pages}{593} (\bibinfo{year}{1963}).

\bibitem[{\citenamefont{McLachlan}(1963)}]{mcl}
\bibinfo{author}{\bibfnamefont{A.~D.} \bibnamefont{McLachlan}},
  \bibinfo{journal}{Proc. R. Soc. A} \textbf{\bibinfo{volume}{271}},
  \bibinfo{pages}{387} (\bibinfo{year}{1963}).

\bibitem[{\citenamefont{Boyer}(1972)}]{boyer}
\bibinfo{author}{\bibfnamefont{T.~H.} \bibnamefont{Boyer}},
  \bibinfo{journal}{Phys. Rev. A} \textbf{\bibinfo{volume}{5}},
  \bibinfo{pages}{1799} (\bibinfo{year}{1972}).

\bibitem[{\citenamefont{Zaremba and Kohn}(1976)}]{kohn}
\bibinfo{author}{\bibfnamefont{E.}~\bibnamefont{Zaremba}} \bibnamefont{and}
  \bibinfo{author}{\bibfnamefont{W.}~\bibnamefont{Kohn}},
  \bibinfo{journal}{Phys. Rev. B} \textbf{\bibinfo{volume}{13}},
  \bibinfo{pages}{2270} (\bibinfo{year}{1976}).

\bibitem[{\citenamefont{Song et~al.}(2012)\citenamefont{Song, Sun, Wang, Jiang,
  Wang, He, Chen, Zhang, Ma, and Xue}}]{song}
\bibinfo{author}{\bibfnamefont{C.-L.} \bibnamefont{Song}},
  \bibinfo{author}{\bibfnamefont{B.}~\bibnamefont{Sun}},
  \bibinfo{author}{\bibfnamefont{Y.-L.} \bibnamefont{Wang}},
  \bibinfo{author}{\bibfnamefont{Y.-P.} \bibnamefont{Jiang}},
  \bibinfo{author}{\bibfnamefont{L.}~\bibnamefont{Wang}},
  \bibinfo{author}{\bibfnamefont{K.}~\bibnamefont{He}},
  \bibinfo{author}{\bibfnamefont{X.}~\bibnamefont{Chen}},
  \bibinfo{author}{\bibfnamefont{P.}~\bibnamefont{Zhang}},
  \bibinfo{author}{\bibfnamefont{X.-C.} \bibnamefont{Ma}}, \bibnamefont{and}
  \bibinfo{author}{\bibfnamefont{Q.-K.} \bibnamefont{Xue}},
  \bibinfo{journal}{Phys. Rev. Lett.} \textbf{\bibinfo{volume}{108}},
  \bibinfo{pages}{156803} (\bibinfo{year}{2012}).

\bibitem[{\citenamefont{Antezza et~al.}(2004)\citenamefont{Antezza, Pitaevskii,
  and Stringari}}]{ant}
\bibinfo{author}{\bibfnamefont{M.}~\bibnamefont{Antezza}},
  \bibinfo{author}{\bibfnamefont{L.~P.} \bibnamefont{Pitaevskii}},
  \bibnamefont{and}
  \bibinfo{author}{\bibfnamefont{S.}~\bibnamefont{Stringari}},
  \bibinfo{journal}{Phys. Rev. A} \textbf{\bibinfo{volume}{70}},
  \bibinfo{pages}{053619} (\bibinfo{year}{2004}).

\bibitem[{\citenamefont{Roberts et~al.}(1998)\citenamefont{Roberts, Claussen,
  Burke, Greene, Cornell, and Wieman}}]{robert}
\bibinfo{author}{\bibfnamefont{J.~L.} \bibnamefont{Roberts}},
  \bibinfo{author}{\bibfnamefont{N.~R.} \bibnamefont{Claussen}},
  \bibinfo{author}{\bibfnamefont{J.~P.} \bibnamefont{Burke}},
  \bibinfo{author}{\bibfnamefont{C.~H.} \bibnamefont{Greene}},
  \bibinfo{author}{\bibfnamefont{E.~A.} \bibnamefont{Cornell}},
  \bibnamefont{and} \bibinfo{author}{\bibfnamefont{C.~E.}
  \bibnamefont{Wieman}}, \bibinfo{journal}{Phys. Rev. Lett.}
  \textbf{\bibinfo{volume}{81}}, \bibinfo{pages}{5109} (\bibinfo{year}{1998}).

\bibitem[{\citenamefont{Amiot and Verges}(2000)}]{amiot}
\bibinfo{author}{\bibfnamefont{C.}~\bibnamefont{Amiot}} \bibnamefont{and}
  \bibinfo{author}{\bibfnamefont{J.}~\bibnamefont{Verges}},
  \bibinfo{journal}{J. Chem. Phys.} \textbf{\bibinfo{volume}{112}},
  \bibinfo{pages}{7068} (\bibinfo{year}{2000}).

\bibitem[{\citenamefont{Leo et~al.}(2000)\citenamefont{Leo, Williams, and
  Julienne}}]{leo}
\bibinfo{author}{\bibfnamefont{P.~J.} \bibnamefont{Leo}},
  \bibinfo{author}{\bibfnamefont{C.~J.} \bibnamefont{Williams}},
  \bibnamefont{and} \bibinfo{author}{\bibfnamefont{P.~S.}
  \bibnamefont{Julienne}}, \bibinfo{journal}{Phys. Rev. Lett.}
  \textbf{\bibinfo{volume}{85}}, \bibinfo{pages}{2721} (\bibinfo{year}{2000}).

\bibitem[{\citenamefont{Harber et~al.}(2003)\citenamefont{Harber, McGuirk,
  Obrecht, and Cornell}}]{harber}
\bibinfo{author}{\bibfnamefont{D.~M.} \bibnamefont{Harber}},
  \bibinfo{author}{\bibfnamefont{J.~M.} \bibnamefont{McGuirk}},
  \bibinfo{author}{\bibfnamefont{J.~M.} \bibnamefont{Obrecht}},
  \bibnamefont{and} \bibinfo{author}{\bibfnamefont{E.~A.}
  \bibnamefont{Cornell}}, \bibinfo{journal}{J. Low Temp. Phys.}
  \textbf{\bibinfo{volume}{133}}, \bibinfo{pages}{229} (\bibinfo{year}{2003}).

\bibitem[{\citenamefont{Leanhardt et~al.}(2003)\citenamefont{Leanhardt, Shin,
  Chikkatur, Kielpinski, Ketterle, and Pritchard}}]{lean}
\bibinfo{author}{\bibfnamefont{A.~E.} \bibnamefont{Leanhardt}},
  \bibinfo{author}{\bibfnamefont{Y.}~\bibnamefont{Shin}},
  \bibinfo{author}{\bibfnamefont{A.~P.} \bibnamefont{Chikkatur}},
  \bibinfo{author}{\bibfnamefont{D.}~\bibnamefont{Kielpinski}},
  \bibinfo{author}{\bibfnamefont{W.}~\bibnamefont{Ketterle}}, \bibnamefont{and}
  \bibinfo{author}{\bibfnamefont{D.~E.} \bibnamefont{Pritchard}},
  \bibinfo{journal}{Phys. Rev. Lett.} \textbf{\bibinfo{volume}{90}},
  \bibinfo{pages}{100404} (\bibinfo{year}{2003}).

\bibitem[{\citenamefont{Lin et~al.}(2004)\citenamefont{Lin, Teper, Chin, and
  Vuletic}}]{lin}
\bibinfo{author}{\bibfnamefont{Y.~J.} \bibnamefont{Lin}},
  \bibinfo{author}{\bibfnamefont{I.}~\bibnamefont{Teper}},
  \bibinfo{author}{\bibfnamefont{C.}~\bibnamefont{Chin}}, \bibnamefont{and}
  \bibinfo{author}{\bibfnamefont{V.}~\bibnamefont{Vuletic}},
  \bibinfo{journal}{Phys. Rev. Lett.} \textbf{\bibinfo{volume}{92}},
  \bibinfo{pages}{050404} (\bibinfo{year}{2004}).

\bibitem[{\citenamefont{Arora and Sahoo}(2014)}]{bindiya}
\bibinfo{author}{\bibfnamefont{B.}~\bibnamefont{Arora}} \bibnamefont{and}
  \bibinfo{author}{\bibfnamefont{B.~K.} \bibnamefont{Sahoo}},
  \bibinfo{journal}{Phys. Rev. A} \textbf{\bibinfo{volume}{89}},
  \bibinfo{pages}{022511} (\bibinfo{year}{2014}).

\bibitem[{\citenamefont{Folman et~al.}(2000)\citenamefont{Folman, Kruger,
  Cassettari, Hessmo, Maier, and Schmiedmayer}}]{folman}
\bibinfo{author}{\bibfnamefont{R.}~\bibnamefont{Folman}},
  \bibinfo{author}{\bibfnamefont{P.}~\bibnamefont{Kruger}},
  \bibinfo{author}{\bibfnamefont{D.}~\bibnamefont{Cassettari}},
  \bibinfo{author}{\bibfnamefont{B.}~\bibnamefont{Hessmo}},
  \bibinfo{author}{\bibfnamefont{T.}~\bibnamefont{Maier}}, \bibnamefont{and}
  \bibinfo{author}{\bibfnamefont{J.}~\bibnamefont{Schmiedmayer}},
  \bibinfo{journal}{Phys. Rev. Lett.} \textbf{\bibinfo{volume}{84}},
  \bibinfo{pages}{4749} (\bibinfo{year}{2000}).

\bibitem[{\citenamefont{Kharchenko et~al.}(1997)\citenamefont{Kharchenko, Babb,
  and Dalgarno}}]{kharchenko}
\bibinfo{author}{\bibfnamefont{P.}~\bibnamefont{Kharchenko}},
  \bibinfo{author}{\bibfnamefont{J.~F.} \bibnamefont{Babb}}, \bibnamefont{and}
  \bibinfo{author}{\bibfnamefont{A.}~\bibnamefont{Dalgarno}},
  \bibinfo{journal}{Phys. Rev. A} \textbf{\bibinfo{volume}{55}},
  \bibinfo{pages}{3566} (\bibinfo{year}{1997}).

\bibitem[{\citenamefont{Lach et~al.}(2010)\citenamefont{Lach, Dekieviet, and
  Jentschura}}]{lach1}
\bibinfo{author}{\bibfnamefont{G.}~\bibnamefont{Lach}},
  \bibinfo{author}{\bibfnamefont{M.}~\bibnamefont{Dekieviet}},
  \bibnamefont{and} \bibinfo{author}{\bibfnamefont{U.~D.}
  \bibnamefont{Jentschura}}, \bibinfo{journal}{Int. J. Mod. Phys. A}
  \textbf{\bibinfo{volume}{25}}, \bibinfo{pages}{2337} (\bibinfo{year}{2010}).

\bibitem[{\citenamefont{Caride et~al.}(2005)\citenamefont{Caride,
  Klimchitskaya, Mostepanenko, and Zanette}}]{caride}
\bibinfo{author}{\bibfnamefont{A.~O.} \bibnamefont{Caride}},
  \bibinfo{author}{\bibfnamefont{G.~L.} \bibnamefont{Klimchitskaya}},
  \bibinfo{author}{\bibfnamefont{V.~M.} \bibnamefont{Mostepanenko}},
  \bibnamefont{and} \bibinfo{author}{\bibfnamefont{S.~I.}
  \bibnamefont{Zanette}}, \bibinfo{journal}{Phys. Rev. A}
  \textbf{\bibinfo{volume}{71}}, \bibinfo{pages}{042901}
  (\bibinfo{year}{2005}).

\bibitem[{\citenamefont{Babb et~al.}(2004)\citenamefont{Babb, Klimchitskaya,
  and Mostepanenko}}]{babb1}
\bibinfo{author}{\bibfnamefont{J.~F.} \bibnamefont{Babb}},
  \bibinfo{author}{\bibfnamefont{G.~L.} \bibnamefont{Klimchitskaya}},
  \bibnamefont{and} \bibinfo{author}{\bibfnamefont{V.~M.}
  \bibnamefont{Mostepanenko}}, \bibinfo{journal}{Phys. Rev. A}
  \textbf{\bibinfo{volume}{70}}, \bibinfo{pages}{042901}
  (\bibinfo{year}{2004}).

\bibitem[{\citenamefont{Vidali et~al.}(1991)\citenamefont{Vidali, Ihm, Kim, and
  Cole}}]{vida}
\bibinfo{author}{\bibfnamefont{G.}~\bibnamefont{Vidali}},
  \bibinfo{author}{\bibfnamefont{G.}~\bibnamefont{Ihm}},
  \bibinfo{author}{\bibfnamefont{H.~Y.} \bibnamefont{Kim}}, \bibnamefont{and}
  \bibinfo{author}{\bibfnamefont{M.~W.} \bibnamefont{Cole}},
  \bibinfo{journal}{Surf. Sci. Rep.} \textbf{\bibinfo{volume}{12}},
  \bibinfo{pages}{133} (\bibinfo{year}{1991}).

\bibitem[{\citenamefont{Lonij et~al.}(2010)\citenamefont{Lonij, Klauss,
  Holmgren, and Cronin}}]{lonij1}
\bibinfo{author}{\bibfnamefont{V.~P.~A.} \bibnamefont{Lonij}},
  \bibinfo{author}{\bibfnamefont{C.~E.} \bibnamefont{Klauss}},
  \bibinfo{author}{\bibfnamefont{W.~F.} \bibnamefont{Holmgren}},
  \bibnamefont{and} \bibinfo{author}{\bibfnamefont{A.~D.}
  \bibnamefont{Cronin}}, \bibinfo{journal}{Phys. Rev. Lett.}
  \textbf{\bibinfo{volume}{105}}, \bibinfo{pages}{233202}
  (\bibinfo{year}{2010}).

\bibitem[{\citenamefont{Lonij et~al.}(2011)\citenamefont{Lonij, Klauss,
  Holmgren, and Cronin}}]{lonij2}
\bibinfo{author}{\bibfnamefont{V.~P.~A.} \bibnamefont{Lonij}},
  \bibinfo{author}{\bibfnamefont{C.~E.} \bibnamefont{Klauss}},
  \bibinfo{author}{\bibfnamefont{W.~F.} \bibnamefont{Holmgren}},
  \bibnamefont{and} \bibinfo{author}{\bibfnamefont{A.~D.}
  \bibnamefont{Cronin}}, \bibinfo{journal}{J. Phys. Chem. A}
  \textbf{\bibinfo{volume}{115}}, \bibinfo{pages}{7134} (\bibinfo{year}{2011}).

\bibitem[{\citenamefont{Lifshitz and Pitaevskii}(1980)}]{lifshitzbook}
\bibinfo{author}{\bibfnamefont{E.~M.} \bibnamefont{Lifshitz}} \bibnamefont{and}
  \bibinfo{author}{\bibfnamefont{L.~P.} \bibnamefont{Pitaevskii}},
  \emph{\bibinfo{title}{Statistical Physics}} (\bibinfo{publisher}{Pergamon
  Press}, \bibinfo{address}{Oxford}, \bibinfo{year}{1980}).

\bibitem[{\citenamefont{Perreault et~al.}(2005)\citenamefont{Perreault, Cronin,
  and Savas}}]{cronin}
\bibinfo{author}{\bibfnamefont{J.~D.} \bibnamefont{Perreault}},
  \bibinfo{author}{\bibfnamefont{A.~D.} \bibnamefont{Cronin}},
  \bibnamefont{and} \bibinfo{author}{\bibfnamefont{T.~A.} \bibnamefont{Savas}},
  \bibinfo{journal}{Phys. Rev. A} \textbf{\bibinfo{volume}{71}},
  \bibinfo{pages}{053612} (\bibinfo{year}{2005}).

\bibitem[{\citenamefont{Fichet et~al.}(1995)\citenamefont{Fichet, Schuller,
  Bloch, and Ducloy}}]{fichet}
\bibinfo{author}{\bibfnamefont{M.}~\bibnamefont{Fichet}},
  \bibinfo{author}{\bibfnamefont{F.}~\bibnamefont{Schuller}},
  \bibinfo{author}{\bibfnamefont{D.}~\bibnamefont{Bloch}}, \bibnamefont{and}
  \bibinfo{author}{\bibfnamefont{M.}~\bibnamefont{Ducloy}},
  \bibinfo{journal}{Phys. Rev. A} \textbf{\bibinfo{volume}{51}},
  \bibinfo{pages}{1553} (\bibinfo{year}{1995}).

\bibitem[{\citenamefont{Rafac et~al.}(1999)\citenamefont{Rafac, Tanner,
  Livingston, and Berry}}]{rafac}
\bibinfo{author}{\bibfnamefont{R.~J.} \bibnamefont{Rafac}},
  \bibinfo{author}{\bibfnamefont{C.~E.} \bibnamefont{Tanner}},
  \bibinfo{author}{\bibfnamefont{A.~E.} \bibnamefont{Livingston}},
  \bibnamefont{and} \bibinfo{author}{\bibfnamefont{H.~G.} \bibnamefont{Berry}},
  \bibinfo{journal}{Phys. Rev. A} \textbf{\bibinfo{volume}{60}},
  \bibinfo{pages}{3648} (\bibinfo{year}{1999}).

\bibitem[{\citenamefont{Borschevsky et~al.}(2013)\citenamefont{Borschevsky,
  Pershina, Eliav, and Kaldor}}]{borch}
\bibinfo{author}{\bibfnamefont{A.}~\bibnamefont{Borschevsky}},
  \bibinfo{author}{\bibfnamefont{V.}~\bibnamefont{Pershina}},
  \bibinfo{author}{\bibfnamefont{E.}~\bibnamefont{Eliav}}, \bibnamefont{and}
  \bibinfo{author}{\bibfnamefont{U.}~\bibnamefont{Kaldor}},
  \bibinfo{journal}{J. Chem. Phys.} \textbf{\bibinfo{volume}{138}},
  \bibinfo{pages}{124302} (\bibinfo{year}{2013}).

\bibitem[{\citenamefont{Derevianko et~al.}(2010)\citenamefont{Derevianko,
  Porsev, and Babb}}]{dere3}
\bibinfo{author}{\bibfnamefont{A.}~\bibnamefont{Derevianko}},
  \bibinfo{author}{\bibfnamefont{S.~G.} \bibnamefont{Porsev}},
  \bibnamefont{and} \bibinfo{author}{\bibfnamefont{J.~F.} \bibnamefont{Babb}},
  \bibinfo{journal}{Atomic Data and Nuclear Data Tables}
  \textbf{\bibinfo{volume}{96}}, \bibinfo{pages}{323} (\bibinfo{year}{2010}).

\bibitem[{\citenamefont{Amini and Gould}(2003)}]{amini}
\bibinfo{author}{\bibfnamefont{J.~M.} \bibnamefont{Amini}} \bibnamefont{and}
  \bibinfo{author}{\bibfnamefont{H.}~\bibnamefont{Gould}},
  \bibinfo{journal}{Phys. Rev. Lett.} \textbf{\bibinfo{volume}{91}},
  \bibinfo{pages}{153001} (\bibinfo{year}{2003}).

\bibitem[{\citenamefont{Arora et~al.}(2012)\citenamefont{Arora, Nandy, and
  Sahoo}}]{nandy}
\bibinfo{author}{\bibfnamefont{B.}~\bibnamefont{Arora}},
  \bibinfo{author}{\bibfnamefont{D.~K.} \bibnamefont{Nandy}}, \bibnamefont{and}
  \bibinfo{author}{\bibfnamefont{B.~K.} \bibnamefont{Sahoo}},
  \bibinfo{journal}{Phys. Rev. A} \textbf{\bibinfo{volume}{85}},
  \bibinfo{pages}{012506} (\bibinfo{year}{2012}).

\bibitem[{\citenamefont{Kaur et~al.}(2015)\citenamefont{Kaur, Nandy, Arora, and
  Sahoo}}]{jasmeet}
\bibinfo{author}{\bibfnamefont{J.}~\bibnamefont{Kaur}},
  \bibinfo{author}{\bibfnamefont{D.~K.} \bibnamefont{Nandy}},
  \bibinfo{author}{\bibfnamefont{B.}~\bibnamefont{Arora}}, \bibnamefont{and}
  \bibinfo{author}{\bibfnamefont{B.~K.} \bibnamefont{Sahoo}},
  \bibinfo{journal}{Phys. Rev. A} \textbf{\bibinfo{volume}{91}},
  \bibinfo{pages}{012705} (\bibinfo{year}{2015}).

\bibitem[{\citenamefont{Derevianko et~al.}(1998)\citenamefont{Derevianko,
  Johnson, and Fritzsche}}]{dere1}
\bibinfo{author}{\bibfnamefont{A.}~\bibnamefont{Derevianko}},
  \bibinfo{author}{\bibfnamefont{W.~R.} \bibnamefont{Johnson}},
  \bibnamefont{and}
  \bibinfo{author}{\bibfnamefont{S.}~\bibnamefont{Fritzsche}},
  \bibinfo{journal}{Phys. Rev. A} \textbf{\bibinfo{volume}{57}},
  \bibinfo{pages}{2629} (\bibinfo{year}{1998}).

\bibitem[{\citenamefont{Derevianko et~al.}(1999)\citenamefont{Derevianko,
  Johnson, Safronova, and Babb}}]{dere2}
\bibinfo{author}{\bibfnamefont{A.}~\bibnamefont{Derevianko}},
  \bibinfo{author}{\bibfnamefont{W.~R.} \bibnamefont{Johnson}},
  \bibinfo{author}{\bibfnamefont{M.~S.} \bibnamefont{Safronova}},
  \bibnamefont{and} \bibinfo{author}{\bibfnamefont{J.~F.} \bibnamefont{Babb}},
  \bibinfo{journal}{Phys. Rev. Lett.} \textbf{\bibinfo{volume}{82}},
  \bibinfo{pages}{3589} (\bibinfo{year}{1999}).

\bibitem[{\citenamefont{Arora and Sahoo}(2012)}]{bindiya1}
\bibinfo{author}{\bibfnamefont{B.}~\bibnamefont{Arora}} \bibnamefont{and}
  \bibinfo{author}{\bibfnamefont{B.~K.} \bibnamefont{Sahoo}},
  \bibinfo{journal}{Phys. Rev. A} \textbf{\bibinfo{volume}{86}},
  \bibinfo{pages}{033416} (\bibinfo{year}{2012}).

\bibitem[{\citenamefont{Arora et~al.}(2007)\citenamefont{Arora, Safronova, and
  Clark}}]{bindiya2}
\bibinfo{author}{\bibfnamefont{B.}~\bibnamefont{Arora}},
  \bibinfo{author}{\bibfnamefont{M.~S.} \bibnamefont{Safronova}},
  \bibnamefont{and} \bibinfo{author}{\bibfnamefont{C.~W.} \bibnamefont{Clark}},
  \bibinfo{journal}{Phys. Rev. A} \textbf{\bibinfo{volume}{76}},
  \bibinfo{pages}{052509} (\bibinfo{year}{2007}).

\bibitem[{\citenamefont{Sahoo and Das}(2015)}]{bijaya1}
\bibinfo{author}{\bibfnamefont{B.~K.} \bibnamefont{Sahoo}} \bibnamefont{and}
  \bibinfo{author}{\bibfnamefont{B.~P.} \bibnamefont{Das}},
  \bibinfo{journal}{Phys. Rev. A} \textbf{\bibinfo{volume}{92}},
  \bibinfo{pages}{052511} (\bibinfo{year}{2015}).

\bibitem[{\citenamefont{Sahoo}()}]{bijaya2}
\bibinfo{author}{\bibfnamefont{B.~K.} \bibnamefont{Sahoo}},
  \bibinfo{note}{http://arxiv.org/abs/1510.08351}.

\bibitem[{\citenamefont{Singh et~al.}(2013)\citenamefont{Singh, Sahoo, and
  Das}}]{yashpal}
\bibinfo{author}{\bibfnamefont{Y.}~\bibnamefont{Singh}},
  \bibinfo{author}{\bibfnamefont{B.~K.} \bibnamefont{Sahoo}}, \bibnamefont{and}
  \bibinfo{author}{\bibfnamefont{B.~P.} \bibnamefont{Das}},
  \bibinfo{journal}{Phys. Rev. A} \textbf{\bibinfo{volume}{88}},
  \bibinfo{pages}{062504} (\bibinfo{year}{2013}).

\bibitem[{\citenamefont{Palik}(1985)}]{palik}
\bibinfo{author}{\bibfnamefont{E.~D.} \bibnamefont{Palik}},
  \emph{\bibinfo{title}{Handbook of optical constants of solids}}
  (\bibinfo{year}{1985}), \bibinfo{note}{academic Press, San Diego}.

\bibitem[{\citenamefont{Lambrecht and Reynaud}(2000)}]{lamb1}
\bibinfo{author}{\bibfnamefont{A.}~\bibnamefont{Lambrecht}} \bibnamefont{and}
  \bibinfo{author}{\bibfnamefont{S.}~\bibnamefont{Reynaud}},
  \bibinfo{journal}{Eur. Phys. J. D} \textbf{\bibinfo{volume}{8}},
  \bibinfo{pages}{309} (\bibinfo{year}{2000}).

\bibitem[{\citenamefont{Klimchitskaya et~al.}(2000)\citenamefont{Klimchitskaya,
  Mohideen, and Mostepanenko}}]{klim}
\bibinfo{author}{\bibfnamefont{G.~L.} \bibnamefont{Klimchitskaya}},
  \bibinfo{author}{\bibfnamefont{U.}~\bibnamefont{Mohideen}}, \bibnamefont{and}
  \bibinfo{author}{\bibfnamefont{V.~M.} \bibnamefont{Mostepanenko}},
  \bibinfo{journal}{Phys. Rev. A} \textbf{\bibinfo{volume}{61}},
  \bibinfo{pages}{062107} (\bibinfo{year}{2000}).

\bibitem[{\citenamefont{Lambrecht et~al.}(2007)\citenamefont{Lambrecht,
  Pirozhenko, Duraffourg, and Andreucci}}]{lamb}
\bibinfo{author}{\bibfnamefont{A.}~\bibnamefont{Lambrecht}},
  \bibinfo{author}{\bibfnamefont{I.}~\bibnamefont{Pirozhenko}},
  \bibinfo{author}{\bibfnamefont{L.}~\bibnamefont{Duraffourg}},
  \bibnamefont{and}
  \bibinfo{author}{\bibfnamefont{P.}~\bibnamefont{Andreucci}},
  \bibinfo{journal}{Eur. Phys. Lett.} \textbf{\bibinfo{volume}{77}},
  \bibinfo{pages}{44006} (\bibinfo{year}{2007}).

\bibitem[{lin()}]{link}
\bibinfo{note}{See http://www.wolfram.com/mathematica.}

\bibitem[{\citenamefont{Kiraha and Honda}(1965)}]{kiraha}
\bibinfo{author}{\bibfnamefont{T.}~\bibnamefont{Kiraha}} \bibnamefont{and}
  \bibinfo{author}{\bibfnamefont{N.}~\bibnamefont{Honda}}, \bibinfo{journal}{J.
  Phys. Soc. Jpn} \textbf{\bibinfo{volume}{20}}, \bibinfo{pages}{15}
  (\bibinfo{year}{1965}).

\bibitem[{\citenamefont{Bruch and Watanabe}(1977)}]{bruch}
\bibinfo{author}{\bibfnamefont{L.~W.} \bibnamefont{Bruch}} \bibnamefont{and}
  \bibinfo{author}{\bibfnamefont{H.}~\bibnamefont{Watanabe}},
  \bibinfo{journal}{Surf. Sci.} \textbf{\bibinfo{volume}{65}},
  \bibinfo{pages}{619} (\bibinfo{year}{1977}).

\bibitem[{\citenamefont{Laliotis and Ducloy}(2015)}]{lali}
\bibinfo{author}{\bibfnamefont{A.}~\bibnamefont{Laliotis}} \bibnamefont{and}
  \bibinfo{author}{\bibfnamefont{M.}~\bibnamefont{Ducloy}},
  \bibinfo{journal}{Phys. Rev. A} \textbf{\bibinfo{volume}{91}},
  \bibinfo{pages}{052506} (\bibinfo{year}{2015}).

\bibitem[{\citenamefont{Foster et~al.}(2004)\citenamefont{Foster, Gal, Gale,
  Lee, Nieminen, and Shluger}}]{foster}
\bibinfo{author}{\bibfnamefont{A.~S.} \bibnamefont{Foster}},
  \bibinfo{author}{\bibfnamefont{A.~Y.} \bibnamefont{Gal}},
  \bibinfo{author}{\bibfnamefont{J.~D.} \bibnamefont{Gale}},
  \bibinfo{author}{\bibfnamefont{Y.~J.} \bibnamefont{Lee}},
  \bibinfo{author}{\bibfnamefont{R.~M.} \bibnamefont{Nieminen}},
  \bibnamefont{and} \bibinfo{author}{\bibfnamefont{A.~L.}
  \bibnamefont{Shluger}}, \bibinfo{journal}{Phy. Rev. Lett.}
  \textbf{\bibinfo{volume}{92}}, \bibinfo{pages}{036101}
  (\bibinfo{year}{2004}).

\end{thebibliography}

\end{document}